\documentclass{osa-article}

\journal{oe}

\begin{document}

\title{Flat liquid jet as a highly efficient source of terahertz radiation}

\author{Anton N. Tcypkin,\authormark{1,*} Evgenia A. Ponomareva,\authormark{1} Sergey E. Putilin,\authormark{1} Semen V. Smirnov \authormark{1,2} Sviatoslav A. Shtumpf,\authormark{1} Maksim V. Melnik,\authormark{1} Yiwen E,\authormark{3,4} Sergei A. Kozlov,\authormark{1} and Xi-Cheng Zhang\authormark{1,3,4}}

\address{\authormark{1}International Laboratory of Femtosecond Optics and Femtotechnologies, ITMO University, St. Petersburg, 197101, Russia\\
\authormark{2} Aston Institute of Photonic Technologies, Aston Univeristy, B4 7ET, Birmingham, UK\\
\authormark{3}The Institute of Optics, University of Rochester, Rochester, NY 14627, USA\\
\authormark{4}Beijing Advanced Innovation Center for Imaging Technology, Capital Normal University, Beijing 100037, China}

\email{\authormark{*}tsypkinan@corp.ifmo.ru} 



\begin{abstract}
Polar liquids are strong absorbers of electromagnetic waves in the terahertz range, therefore, historically such liquids have not been considered as good candidates for terahertz sources. However, flowing liquid medium has explicit advantages, such as a higher damage threshold compared to solid-state sources and more efficient ionization process compared to gases. Here we report systematic study of efficient generation of terahertz radiation in flat liquid jets under sub-picosecond single-color optical excitation. We demonstrate how medium parameters such as molecular density, ionization energy and linear absorption contribute to the terahertz emission from the flat liquid jets. Our simulation and experimental measurements reveal that the terahertz energy has quasi-quadratic dependence on the optical excitation pulse energy. Moreover, the optimal pump pulse duration, which depends on the thickness of the jet is theoretically predicted and experimentally confirmed. The obtained optical-to-terahertz energy conversion efficiency is more than 0.05\%. It is comparable to the commonly used optical rectification in most of electro-optical crystals and two-color air filamentation. These results, significantly advancing prior research, can be successfully applied to create a new alternative source of terahertz radiation.
\end{abstract}

\section{Introduction}
Terahertz wave generation is one of hot research topics for the last decade. Broadband terahertz radiation is suitable for a wide range of applications, for example, in nonlinear terahertz optics \cite{zhang2017extreme}, optical excitations \cite{cocker2013ultrafast}, ultrafast dynamics \cite{tonouchi2007cutting, kampfrath2013resonant} and terahertz informational technologies \cite{baierl2016nonlinear}. Current scientific papers demonstrate the possibility of manipulating the parameters of output terahertz radiation such as polarization \cite{zhang2018manipulation} and energy by controlling the input conditions. The main parameter of the terahertz radiation sources is optical-to-terahertz energy conversion efficiency. There are highly efficient methods of terahertz generation, for example, via optical rectification in crystals, which are limited by the destruction of the material with increase in the pump energy above the damage threshold \cite{vicario2014generation, zhang2010introduction, zhukova2018two}. It is worth noting the new variety of materials on the basis of which the efficient sources of terahertz radiation are being created \cite{wang2018high, esaulkov2015emission}. The most popular mechanism is terahertz generation via filamentation in gases with one or two-color optical excitation due to the possibility to control output radiation by the pump laser parameters \cite{kim2008coherent, kuk2016generation, buccheri2015terahertz, liu2016enhanced}. Using two-color filamentation, maximum efficiency is around 0.01\% \cite{oh2014generation}. The increase of conversion efficiency in the case of filamentation occurs using a different wavelength of the pump radiation \cite{fedorov2018generation}. In this regard, the search for new sources is unlimited and opens a vast area for further investigation.

Until recently, the use of liquids as terahertz sources has been poorly studied, although it certainly has a great potential. Unlike in solid-state sources of terahertz radiation, permanent damage can be avoided in liquids due to their fluidity. Moreover, different nonlinear effects become more perceptible in liquids due to its 3 orders higher molecular density and slightly lower ionization energy compared to gases which results in more charged particles produced in the same ionized volume \cite{williams1976liquid, nikogosyan1983two, crowell1996multiphoton}. The first studies demonstrated the generation of a broadband terahertz waves from a gravitational, free-flowing water film \cite{jin2017observation} and liquids in a cell \cite{dey2017highly} based on plasma filament inside ionized liquid medium. The use of films in previous experiments led to the destruction of the liquid samples at energies above 500 $\mu$J. Undoubtedly, in our case the replacement of the film by a flat jet removes this shortcoming by increasing the liquid damage threshold by more than an order of magnitude. However, it is necessary to make a complete analysis of the physics of terahertz generation in liquid medium where we consider all experimental conditions, namely, the physical and chemical properties of the liquids and the pump radiation parameters. It has not been done yet, although it can significantly advance prior research.

Here, we unravel the secrets that will allow us to improve the terahertz generation in flat liquid jets considering the variety of liquid properties and initial radiation parameters based on theoretical and experimental investigations. Comparing water, heavy water, acetone and ethanol as a generating liquid medium we draw conclusions of the terahertz radiation energy dependence on the duration and the input pulse energy.  We determine that the efficiency of terahertz radiation generation depends on the combined conditions associated with liquid molecular density, its ionization energy, which defines the number of electrons that can be removed from the molecule, and the linear absorption of the liquid. Hence, it follows that all factors must be taken into account. Our experimental data and numerical simulations show that varying the thickness of the medium, we can obtain the maximum efficiency generation of terahertz radiation on different pump duration. We demonstrate that using single-color filamentation in flat liquid jets with increasing pump energy we obtain a quasi-quadratic rise of the output terahertz energy. Finally, by applying this technique we achieve the efficiency of generation of more than 0.05\%, which is comparable with well-known process of two-color filamentation in air. It makes this technique very attractive for a wide range of applications. All these aspects have not been analyzed in detail in previous articles. This paper is another important step towards understanding of the hitherto unexplored physics of terahertz generation in liquids.

\section{Experimental Setup}
\begin{figure}[h!]
\centering
\fbox{\includegraphics[width=\linewidth]{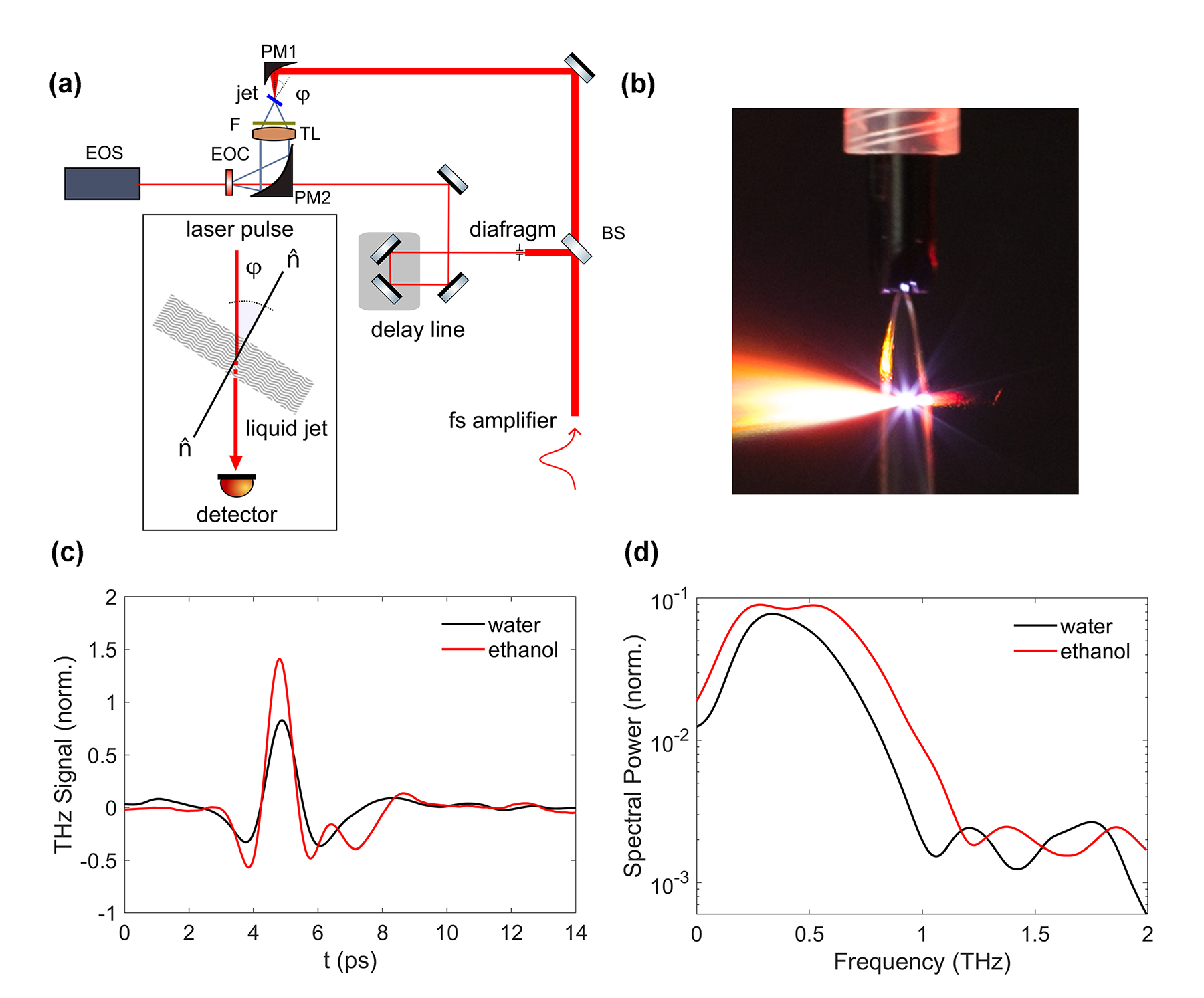}}
\caption{Experimental setup of terahertz generation in flat liquid jets. (a) Experimental layout for energy and spectral terahertz measurements (the inset shows an illustration of optical incident angle $\phi$). Laser radiation is splat on pump and probe beams with beam-splitter (BS) with ratio of energy in the channels 1:49, for probe and pump, respectively. Parabolic mirror (PM1 with focal length equal 5 cm) focus the pump radiation on a liquid jet which leads to the generation of terahertz radiation as a result of filamentation inside ionizing liquid jet. The terahertz radiation is collected and collimated by TPX lens (TL) filtered by a teflon filter (F). For spectrum measurements we use conventional electro-optical system (EOS). Parabolic mirror (PM2 with focal length equal 12 cm) focus the terahertz radiation on the ZnTe crystal (EOC) with 1 mm thickness. (b) Photo of laser excitation of the liquid jet. Water moisture plum scatter the laser beam. Temporal terahertz signals (c) and spectrum (d) emitted from the jets of water and ethanol with a thickness of 150 $\mu$m at laser pulse duration of 400 fs and optical excitation energy of 600 $\mu$J.}
\label{fig1}
\end{figure}

The experimental setup is shown in Fig. \ref{fig1}(a). We use femtosecond laser with wavelength of 800 nm, and varied pulse duration of 35 fs to 700 fs. The change in duration was carried out by changing the distance between the compressor gratings. In this connection, by changing the duration we received a chirped pulse. The pulse energy up to 2 mJ splits into two paths by a beam splitter (BS) (1:49, for probe and pump, respectively). One path functions as a weak optical probe, with the pulses passing through a delay line before reaching the detection electronics (electro-optical system (EOS), we use 1 mm thick ZnTe crystal, which make it possible to detect a signal up to 3 THz). The other path is used to generate terahertz pulses. The pump radiation is focused with a 5 cm focal length parabolic mirror (PM1) on the flat liquid jet (thickness is 100 $\mu$m, 150 $\mu$m or 270 $\mu$m) with beam waist of 100 $\mu$m. The intensity is $I_0$=2$\cdot$10$^{13}$ W/cm$^2$  for 600 $\mu$J energy and pulse duration of 400 fs. The jet thickness is measured using the cross-correlation function. In our experiments, the flat water jet is located in the middle of the filament.

An illustration of incident angle $\phi$ is shown in the inset of Fig. \ref{fig1}(a), where $\hat{n}$ is the surface normal of the jet. The incidence angles of the flat liquid jets are established in accordance with the results of previous experiments \cite{yiwen2018terahertz}. The corresponding value for water and heavy water is 65$^\circ$, acetone and ethanol is 60$^\circ$. 

Fig. \ref{fig1}(b) is pump-medium interaction close-up. The jet is obtained using the nozzle which combines the compressed-tube nozzle and two razor blades \cite{watanabe1989new}. This design forms a flat water surface with a laminar flow. The optical path of the pump pulse passes through the center of the jet area with a constant thickness. Due to the use of the pump water is released under the pressure. The hydroaccumulator in the system of water supply allows to significantly reduce the pulsations associated with the operation of the water pump. The rate of flow (1 m/s) is sufficient for a complete renewal of water in the area of interaction at 1 kHz laser repetition rate. The jet is installed on the translator, which allow it to rotate, changing the angle of incidence $\phi$ on the jet.

Fig. \ref{fig1}(c), and \ref{fig1}(d) demonstrate temporal terahertz signals and spectrum emitted from the jets of water and ethanol with a thickness of 150 $\mu$m at laser pulse duration of 400 fs and optical excitation energy of 600 $\mu$J, respectively.

\section{Energy and pulse measurements}

\begin{figure}[h!]
\centering
\fbox{\includegraphics[width=\linewidth]{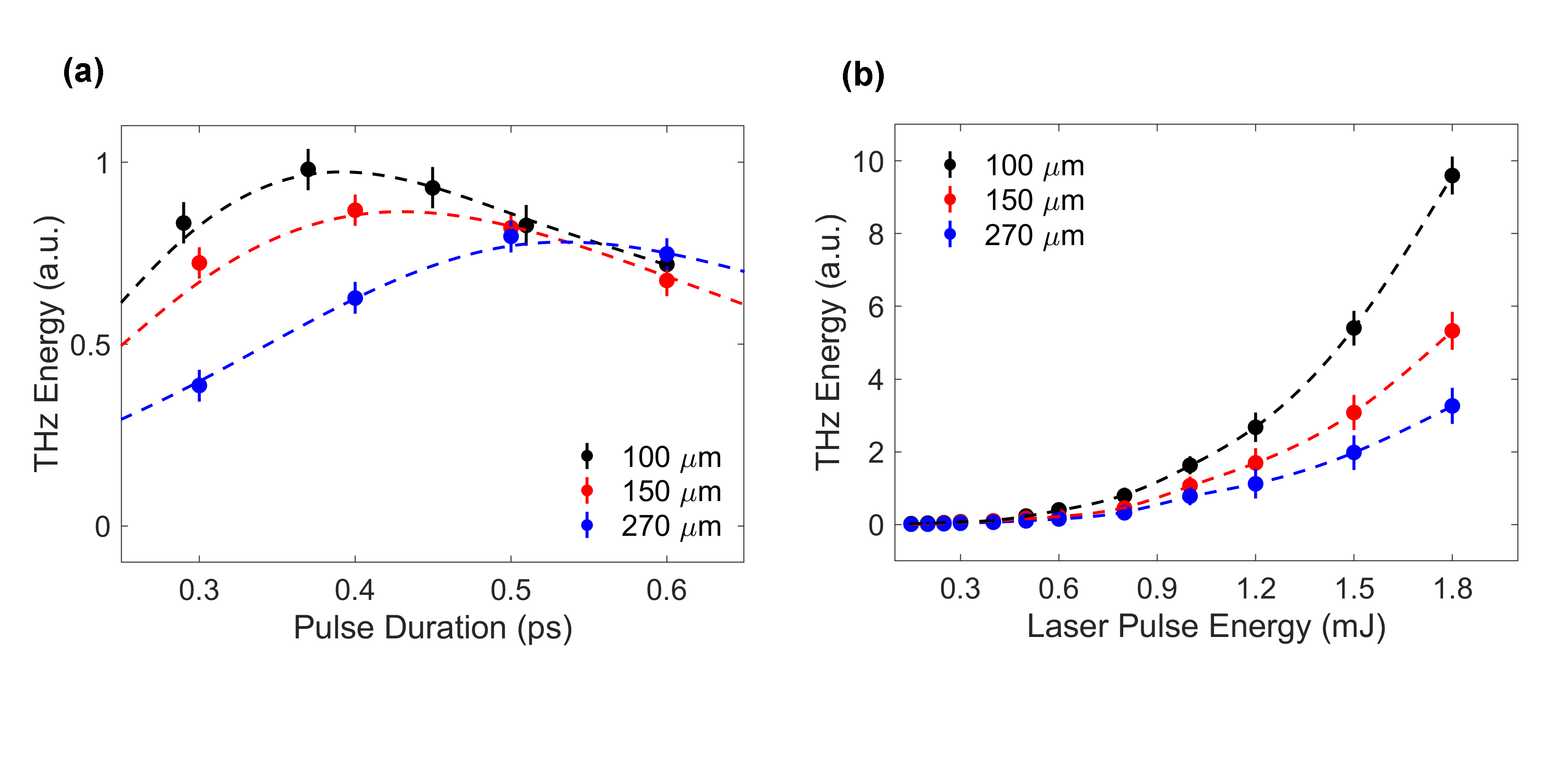}}
\caption{Experimental dependences of the generated terahertz radiation energy characteristics on the energy and duration of the pump pulse for water. Measured terahertz energies generating from single-color filamentation in flat water jet. (a) The dependence of the terahertz energy in flat water jet of different thickness (100 $\mu$m (black), 150 $\mu$m(red) and 270 $\mu$m (blue)) on the laser pulse duration at the pump energy of 600 $\mu$J. (b)  Measured terahertz energies generating from flat water jet as a function of pulse energy with an optimal pulse duration for jet thickness 100 $\mu$m, 150 $\mu$m and 270 $\mu$m. Incidence angle of the flat liquid jet is 65$^\circ$.}
\label{fig2}
\end{figure}

After the first demonstration of the terahertz generation in liquids, the question of how generation efficiency could be increased arose. This ability is necessary for all applications. To begin with, we have carried out experimental studies, where we represent all possible combinations of experimental parameters when generating terahertz radiation in flat liquid jets. We change the parameters of the pump radiation: the duration and energy of the pump pulse, and the thickness of liquid jet. We use water as a reference medium. Our results demonstrate the dependencies of the terahertz generation energy on the pulse duration (Fig. \ref{fig2}(a)) and on the pump energy (Fig. \ref{fig2}(b)) for different thicknesses of flat water jet. We measure the signal energy by integrating the field obtained in the experiment, Fig. \ref{fig1}(c). We note, that the highest recorded values for terahertz radiation energy for a 100 $\mu$m and 150 $\mu$m jet thickness have been obtained for around 350 fs and 400 fs pump pulse, and for 270 $\mu$m jet for around 500 fs. The measurements are performed with a pump energy of 600 $\mu$J. Another interesting point is quasi-quadratic increase in the energy of terahertz radiation with growth of the pump energy (Fig. \ref{fig2}(b)). For a thinner jet of 100 $\mu$m energy is higher. The attenuation of the signal is caused by the increasing absorption as the propagation distance inside the medium increases.

\section{Theoretical model}

The method of theoretical description of the physics of propagation of ultrashort pulses in optical medium, observed in the experiments discussed above, is nontrivial. This method should describe the dynamics of nonlinear processes of medium ionization in the field of a high-intensity ultrashort pulse as well as consider its spectrum superbroadening. 

In the current paper, we calculate the evolution of the optical pulse electric field $E$  itself using equations of the form \cite{stumpf2007few}:

\begin{equation}
\begin{cases}
\frac{\partial E}{\partial z}+\Gamma_0E-a\frac{\partial^3E}{\partial \tau^3}+gE^2\frac{\partial E}{\partial t}+\frac{2\pi}{cn_0}j=0 \\
\frac{\partial j}{\partial \tau}+\frac{j}{\tau_c}=\beta \rho E^3 \\
\frac{\partial \rho}{\partial \tau}+\frac{\rho}{\tau_p}=\alpha E^2
\label{eq1}
\end{cases}
\end{equation}
where  $\Gamma_0$, $n_0$  and $\alpha$  are empirical coefficients characterizing the dependencies of the medium absorption coefficient $\kappa(\omega)=c\Gamma_0/\omega$ and its refractive index \cite{kozlov2013fundamentals} $n(\omega)=n_0+ca\omega^2$ on the frequency $\omega$ which are suitable for describing dispersion in liquids up to 1.2 $\mu$m, since the main part of the radiation is located in this range; $g$  characterizes low-inertia cubic nonlinear response of the medium and connected to its coefficients of the nonlinear refractive index $n_2$ by the ratio \cite{kozlov2013fundamentals} $g=2n_2/c$; $j$  is a current density of quasi-free plasma electrons induced by a strong radiation field; $\rho$ is excited states population of the medium molecules, the transition to which is allowed from their ground state and from which there is a further transition to a quasi-free state under the influence of a field; $\tau_c$ and $\tau_p$ are times of relaxation from quasi-free and excited states, varying which we observe negligible effect on terahertz generation and therefore assumed equal to 1 ps; the coefficients $\alpha$ and $\beta$, characterizing efficiency of the transitions to these states, are fitting parameters for the results of one experiment and will be described in detail below; $z$ is the direction of propagation, $\tau=t-zn_0/c$ is time in the moving frame of reference, $t$ is time, $c$ is the speed of light in vacuum.

Equation (\ref{eq1}) should be supplemented by boundary conditions at  $z$ = 0. We assume that the radiation field at the input surface is yielded by the chirped gaussian pulse \cite{mazurenko1996ultrafast}:

\begin{equation}
E(\tau)= E_0exp \left (-\frac{\tau^2}{\tau_{p}^{2}} \right)sin \left (\omega_0\tau+\frac{A}{\tau_p}(\omega_0\tau)^2 \right)
\label{eq2}
\end{equation}
where $E_0$ is a peak amplitude of the pulse at the input surface, $\omega_0$ is central frequency of radiation with central wavelength $\lambda_0$=800 nm, $A$  is a fitting parameter, representing a pulse chirp, that is chosen such that the width of the chirped pulse spectrum matches width of the  spectrum for the 35 fs spectral-limited pulse, and $\tau_p$ is pulse duration. The intensity value is chosen on the basis of experimental data.

In the numerical calculation the equations (\ref{eq1}) are normalized \cite{andreev2009generation}, and the dynamics of plasma formation can be evaluated by  $(\alpha \cdot \beta)_{liq}$. This value was estimated for water by comparing theoretical and experimental results. For other liquids, it was calculated using the formula $(\alpha \cdot \beta)_{liq}=\frac{\rho_{liq}}{\rho_{water}}\frac{E_{water}^{ion}}{E_{liq}^{ion}}\cdot(\alpha \cdot \beta)_{water}$, where $\rho_{liq}$ is molecular density, $E_{liq}^{ion}$ is ionization energy of the medium.

We must also take into account the linear absorption of the medium, which substantially reduces generation during the propagation through the liquid, by multiplying the field by $exp \left ( -\kappa^{THz}L\right)$, where $\kappa^{THz}$ is linear terahertz absorption of the medium and $L$ is its length.
For the numerical analysis of experimental results, we need estimation of the values of liquid medium parameters (Table \ref{table1}). Numerical simulations were performed for water, heavy water, acetone and ethanol.
\begin{table*}[htbp]
\centering
\caption{\bf Numerical values of the parameters used for simulating the propagation of ultrashort pulse in liquid}
\begin{tabular}{|p{0.12\linewidth}|p{0.12\linewidth}|p{0.12\linewidth}|p{0.12\linewidth}|p{0.12\linewidth}|p{0.12\linewidth}|p{0.12\linewidth}|}
\hline
 & Ionization energy $E_{ion}$, eV & Molecular density \cite{weber2002handbook} $\rho_D$, kg/m$^3$ & THz absorption $\kappa^{THz}$, cm$^{-1}$ & $n_0$ \cite{polyanskiy2016refractive} & $a\cdot$10$^{-44}$, s$^3$/cm & $n_2\cdot$10$^{-16}$, cm$^2$/W\\
\hline
water & 6.5-10.96 \cite{faubel1997photoelectron}, 9 \cite{huang1977effect} & $997$ &$200$ \cite{george2012terahertz} &$1.32$ &$3.68$ &$4.1$ \cite{boyd2008nonlinear} \\
\hline
heavy water & $12.64$ \cite{pang2014water} & $1104$ &$100-200$ \cite{chong2009terahertz} &$1.3184$ &$3.10$ &$6.4$ \cite{weber2002handbook} \\
\hline
aceton & $8.7$ \cite{traeger1982heat} & $784$ &$60$ \cite{venables2000structure} &$1.34979$ &$3.08$ &$24$ \cite{boyd2008nonlinear} \\
\hline
ethanol & $9.7$ \cite{faubel1997photoelectron} & $787$ &$60$ \cite{boyd2008nonlinear} &$1.35265$ &$2.95$ &$7.7$ \cite{boyd2008nonlinear} \\
\hline
\end{tabular}
  \label{table1}
\end{table*}

\section{Discussion}

\begin{figure}[h!]
\centering
\fbox{\includegraphics[width=\linewidth]{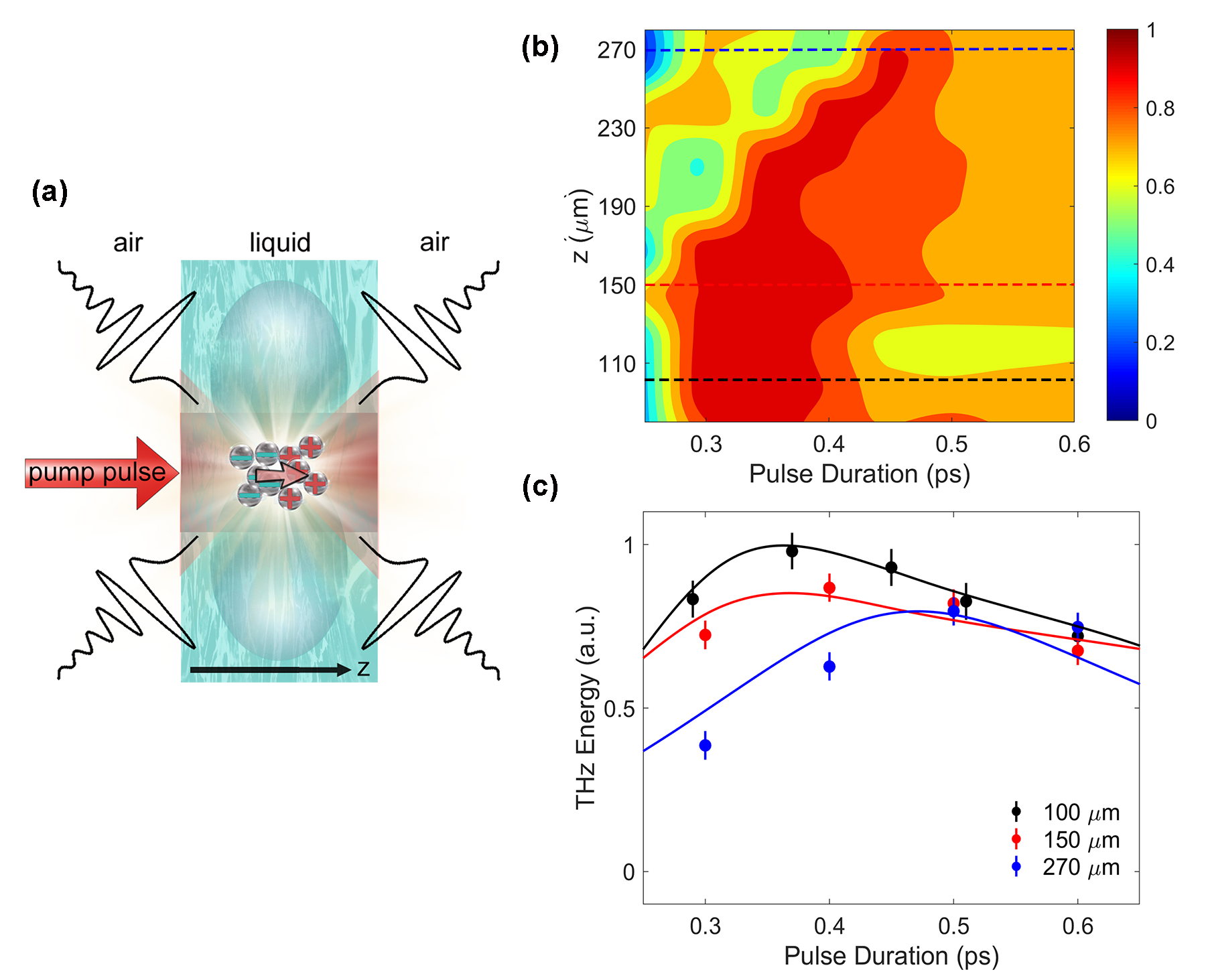}}
\caption{Simulation of terahertz generation in water and its comparison with the experimental data. (a) Visualization of the model description. Here laser pump produces filament with further ionization of the medium and charge separation leading to the terahertz generation. (b) The numerical simulation of the terahertz energy dependence on propagation distance in flat water jet for various pump durations. (c) The comparison of experiment results (dots) and simulation dependencies (lines) of the terahertz energy in flat water jet of different thickness (100 $\mu$m (black), 150 $\mu$m (red) and 270 $\mu$m (blue)) on the laser pulse duration for the pump energy of 600 $\mu$J.}
\label{fig3}
\end{figure}

We use numerical simulation to fit the experimental conditions. Using theoretical model (\ref{eq1}) we simulate terahertz generation in liquid with different input parameters for investigating the theory of the process under study. Visualization of the model description is shown in Fig. \ref{fig3}(a). The propagation of pump radiation in a liquid jet leads to ionization of the medium. The plasma channel is formed and generates bidirectional terahertz waves. Below we present the results for the unidirectional distribution. To obtain the desired range, all frequencies outside 0 to 3 THz is removed from the spectrum for comparison with experiment.

An interesting simulation result is presented in Fig. \ref{fig3}(b). It was previously shown that there is a maximum generation efficiency of terahertz radiation for a certain pump pulse duration. Fig. \ref{fig3}(b) demonstrate that with an increase in the thickness of the medium, the optimal pulse duration increases. The comparison of the experimental results and simulation dependence of the terahertz energy in flat water jets of different thickness (100 $\mu$m (black), 150 $\mu$m (red) and 270 $\mu$m (blue)) on the laser pulse duration at the pump energy of 600 $\mu$J is shown on the Fig. \ref{fig3}(c).

It is known that the pump-medium interaction length is critical for liquids and determined by filament length \cite{dey2017highly}. Using a 5 cm parabolic mirror, the length of the filament is about 1 mm, which is larger than the flat jets thickness in our experiments. Assuming this it is clear that the dominant factor for thin jets is rather a pulse duration than a filament length. With the increase in pulse duration there is a corresponding increase in the interaction time area between the radiation and the medium. Accordingly, the optimal pulse duration should has spatial dimensions greater than or equal to the medium thickness. Due to the more efficient filling of the liquid volume with radiation, there is a growth in number of ionized particles and, accordingly, higher terahertz generation. At the same time, the increase in the pulse duration leads to a decrease in the peak power and deterioration of the electron density \cite{zhang2018optimum}. Accordingly, it follows from the above the existence of the duration local maximum.  

\begin{figure}[h!]
\centering
\fbox{\includegraphics[width=\linewidth]{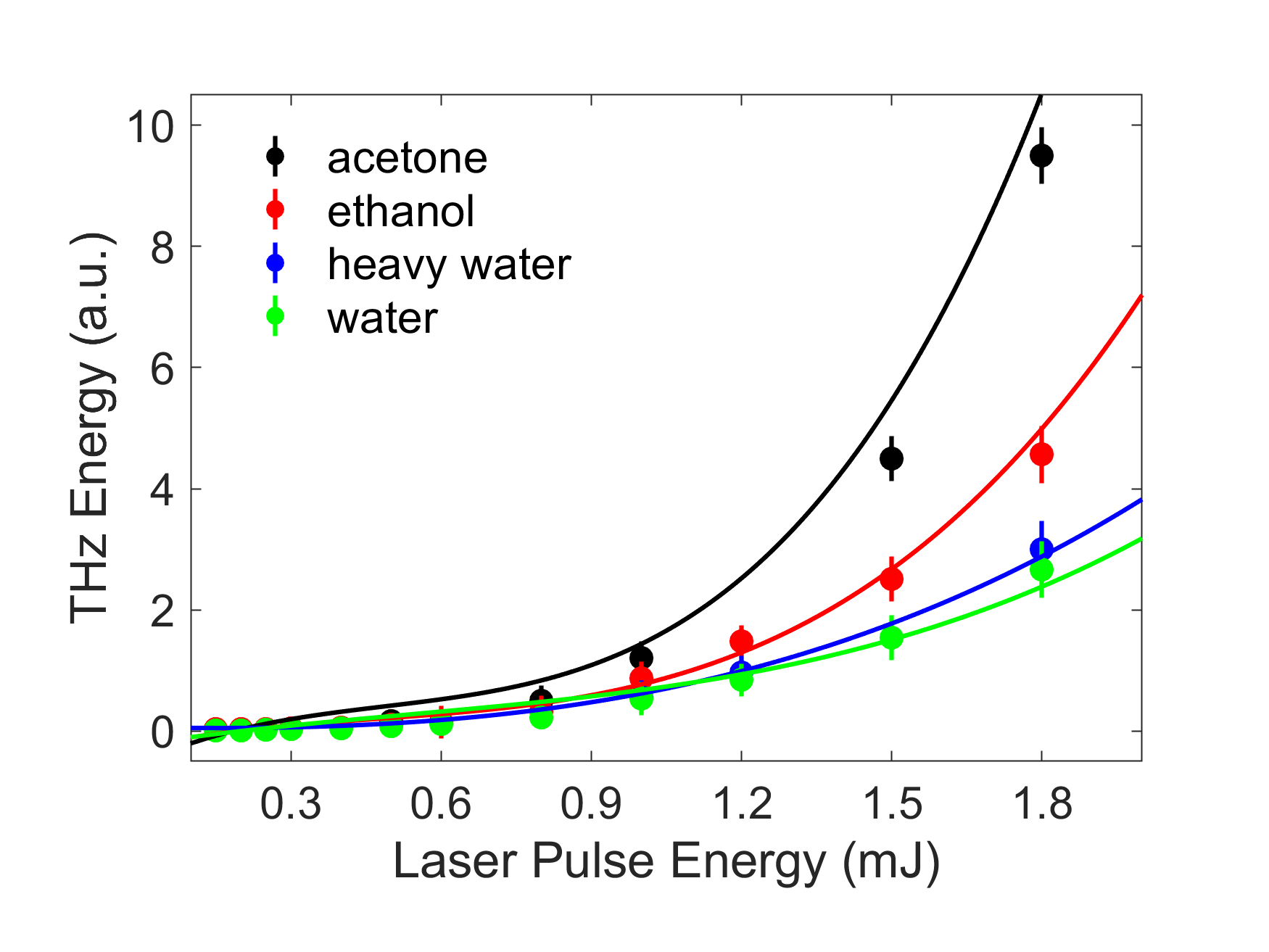}}
\caption{Comparison of experiment and simulation results for the jets of various liquids. Comparison of the numerical simulation result (solid line) of the output terahertz energy dependence on the pump energy with experimental data (scatter) in acetone (black), ethanol (red), heavy water (blue) and water (green) with thickness of 150 $\mu$m.}
\label{fig4}
\end{figure}

In order to find the optimal medium for terahertz generation, we need a comprehensive comparison of their properties. For comparison, we take water, heavy water, acetone and ethanol. Numerical simulation is carried out for comparison with the experimental data and shown in Fig. \ref{fig4}.  We obtain a good agreement between the theory and experiments for all types of liquids. The results confirm the existence of the quasi-quadratic increase in pump energy. Based on the simulation results, we demonstrate the efficiency of terahertz generation depends on the coefficient  $(\alpha \cdot \beta)_{liq}$, which is determined by ionization potential of the medium, its molecular density, and medium absorption in the terahertz range as well. Under the conditions of approximately equal liquid densities, acetone turned out to be the most effective, since this liquid has the lowest absorption in the terahertz range and lowest ionization energy. Thus, considering the entire set of parameters of the medium, it is possible to determine the most advantageous option for creating a terahertz radiation source. 

For the terahertz wave energy measurement in our experiments we compared terahertz field registered on the electro-optical system for two-color filamentation and one color filamentation in flat liquid jets. In case of two color filamentation, we use a spectrally limited pulse with a duration of 35 fs, and as for a liquid jet the optimal pulse for a jet thickness of 100 $\mu$m is 400 fs. We measure the signal energy by integrating the field obtained in the experiment. The results show that the energy of the signals in both cases is comparable. For the terahertz wave energy measurements in our experiments we used commercial Golay cell \cite{yiwen2018terahertz}. For filtering strong mid-IR  radiation emitted under intense optical excitation high-density polyethylene (HDPE) was used. For the input energy of 1.2 mJ, the terahertz energy is 140 nJ for flat ethanol jet of 100 $\mu$m thickness.

\section{Conclusion}

In conclusion, we would like to note that we have highlighted the main experimental conditions that, in our opinion, have significantly contributed to the effective generation of terahertz radiation in flat liquid jets. We developed a mathematical model for the generation of terahertz radiation in liquid jets which predicts the results of experiments. We claim that it is necessary to take into account not only the ionization energy of the medium, but also its molecular density and absorption coefficient to obtain the maximum efficiency of terahertz generation. Comparing all selected liquids, acetone shows the highest efficiency of terahertz generation due to the optimal ratio of all parameters of the medium. We show that it is necessary to consider the correlation between the thickness of the jet and the duration of the pump pulse to obtain the most efficient terahertz generation. Quasi-quadratic dependence of the terahertz energy on the pump pulse energy is demonstrated.  We highlight the achievement of efficiencies of more than 0.05\%, which is comparable with the same value in some solids and more than one magnitude higher than in the air. From the point of view of applications, we believe that our demonstration opens new perspectives for the use of terahertz in the various fields of science and everyday life.

\section*{Funding}

Government of the Russian Federation (08-08); Ministry of Education and Science of the Russian Federation (Minobrnauka) (3.9041.2017/7.8); Army Research Office (W911NF-17-1-0428).



\bibliographystyle{osajnl}
\bibliography{sample}






\end{document}